\documentclass[twocolumn,showpacs,preprintnumbers,amsmath,amssymb,superscriptaddress]{revtex4}

\usepackage{graphicx}
\usepackage{dcolumn}
\usepackage{bm}
\usepackage{graphicx}
\usepackage{xspace}
\usepackage{multirow}
\usepackage{natbib}
\usepackage{color}

\newcommand{\BFA}{BaFe$_2$As$_2$}
\newcommand{\SFA}{SrFe$_2$As$_2$}
\newcommand{\CFA}{CaFe$_2$As$_2$}

\begin{document}

\preprint{}

\title{Observation of Dirac Cone Electronic Dispersion in BaFe$_2$As$_2$}

\author{P. Richard}\email{p.richard@arpes.phys.tohoku.ac.jp}
\affiliation{WPI Research Center, Advanced Institute for Materials Research, Tohoku University, Sendai 980-8577, Japan}
\author{K. Nakayama}
\affiliation{Department of Physics, Tohoku University, Sendai 980-8578, Japan}
\author{T. Sato}
\affiliation{Department of Physics, Tohoku University, Sendai 980-8578, Japan}
\affiliation{TRiP, Japan Science and Technology Agency (JST), Kawaguchi 332-0012, Japan}
\author{M. Neupane}
\affiliation{Department of Physics, Boston College, Chestnut Hill, MA 02467, USA}
\author{Y.-M. Xu}
\affiliation{Department of Physics, Boston College, Chestnut Hill, MA 02467, USA}
\author{J. H. Bowen}
\affiliation{Beijing National Laboratory for Condensed Matter Physics, and Institute of Physics, Chinese Academy of Sciences, Beijing 100190, China}
\author{G. F. Chen}
\affiliation{Beijing National Laboratory for Condensed Matter Physics, and Institute of Physics, Chinese Academy of Sciences, Beijing 100190, China}
\author{J. L. Luo}
\affiliation{Beijing National Laboratory for Condensed Matter Physics, and Institute of Physics, Chinese Academy of Sciences, Beijing 100190, China}
\author{N. L. Wang}
\affiliation{Beijing National Laboratory for Condensed Matter Physics, and Institute of Physics, Chinese Academy of Sciences, Beijing 100190, China}
\author{H. Ding}
\affiliation{Beijing National Laboratory for Condensed Matter Physics, and Institute of Physics, Chinese Academy of Sciences, Beijing 100190, China}
\author{T. Takahashi}
\affiliation{WPI Research Center, Advanced Institute for Materials Research, Tohoku University, Sendai 980-8577, Japan}
\affiliation{Department of Physics, Tohoku University, Sendai 980-8578, Japan}

\date{\today}

\begin{abstract}
We performed an angle-resolved photoemission spectroscopy study of BaFe$_2$As$_2$, which is the parent compound of the so-called \emph{122} phase of the iron-pnictide high-temperature superconductors. We reveal the existence of a Dirac cone in the electronic structure of this material below the spin-density-wave temperature, which is responsible for small spots of high photoemission intensity at the Fermi level. Our analysis suggests that the cone is slightly anisotropic and its apex is located very near the Fermi level, leading to tiny Fermi surface pockets. Moreover, the bands forming the cone show an anisotropic leading edge gap away from the cone that suggests a nodal spin-density-wave description.
\end{abstract}


\pacs{74.70.-b, 74.25.Jb, 79.60.-i,71.18.+y,}
\keywords{Ferropnictides, ARPES, kink \sep HTSC
}
\maketitle


As with cuprates, it is widely believed that high-$T_c$ superconductivity in pnictides emerges by tuning interactions already present in the parent compounds. For example, high-$T_c$ superconductivity up to 37 K is achieved in metallic and antiferromagnetic \BFA\xspace by adding carriers \cite{GF_Chen2, Rotter} or applying pressure \cite{Kimber, Torikachvili2}. Although long-range magnetic order is suppressed at optimal $T_c$, previous angle-resolved photoemission spectroscopy (ARPES)\cite{Ding_EPL, RichardPRL2009} and neutron scattering studies \cite{Christianson} strongly suggest the importance of magnetic fluctuations for the pairing mechanism. While it is admitted that the parent compounds \BFA, \SFA\xspace and \CFA\xspace exhibit a magnetic ordering below a transition temperature $T_{SDW}$ that is accompanied by an orthorhombic distortion \cite{Huang, Zhao, Wilson}, the nature of the force driving this transition remains controversial. 

Owing to its angle-resolved capability to measure directly single-particle electronic spectra, ARPES is a powerful tool that allows a precise description of the Fermi surface (FS) topology and the electronic states lying in the close vicinity of the Fermi level ($E_F$). However, previous attempts by ARPES to elucidate the nature of the low temperature electronic states of the pnictide parent compounds support competing scenarios, including spin-density-wave (SDW) \cite{Hsieh} and exchange splitting \cite{Yang_LX} models. Even more challenging is to answer weather the signatures of tiny FSs inferred from quantum oscillation experiments (QOE)  \cite{Sebastian,Analytis}, which are unfortunately not experimentally located in the momentum space, are analogous to the FS reconstructions claimed in cuprates \cite{Doyron_Leyraud} or consequences of more fundamental topological anomalies in the electronic band structure, such as the Dirac cone in graphene \cite{Novoselov}. It is thus imperative to characterize directly and more precisely the very electronic states that trigger the low temperature properties of the pnictide parent compounds.

In this Letter, we report an ARPES investigation of the low-energy electronic states of \BFA. We show that low temperature \BFA\xspace is better described as a nodal SDW material \cite{Ran}. We reveal the existence of an anisotropic Dirac cone located away from high symmetry points and formed by bands that are gapped away from the cone. The cone's apex is located very near $E_F$, which implies tiny FS pockets.


High-quality single crystals of \BFA\xspace($T_{SDW}$ = 138 K) have been grown using the flux method \cite{GF_Chen2}. A microwave-driven helium source ($h\nu$ = 21.218 eV) and a VG-Scienta SES 2002 multi-channel analyzer were used to record ARPES spectra with energy and angular resolutions of 7-14 meV and 0.2$^{\circ}$, respectively. The samples were cleaved \emph{in-situ} at 25 K and measured with a working vacuum better than 5$\times$10$^{-11}$ Torr within the 25 K to 170 K temperature range. No obvious degradation of the spectra was observed for typical measurements of 3 days. A freshly evaporated gold sample in electrical contact with the \BFA\xspace sample served to calibrate $E_F$. To facilitate data representation, we describe all the results in terms of the unreconstructed Brillouin zone (BZ) formed by the Fe network alone, with the lattice parameter $a$ representing the distance between Fe atoms.

\begin{figure}[htbp]
\begin{center}
\includegraphics[width=8.8cm]{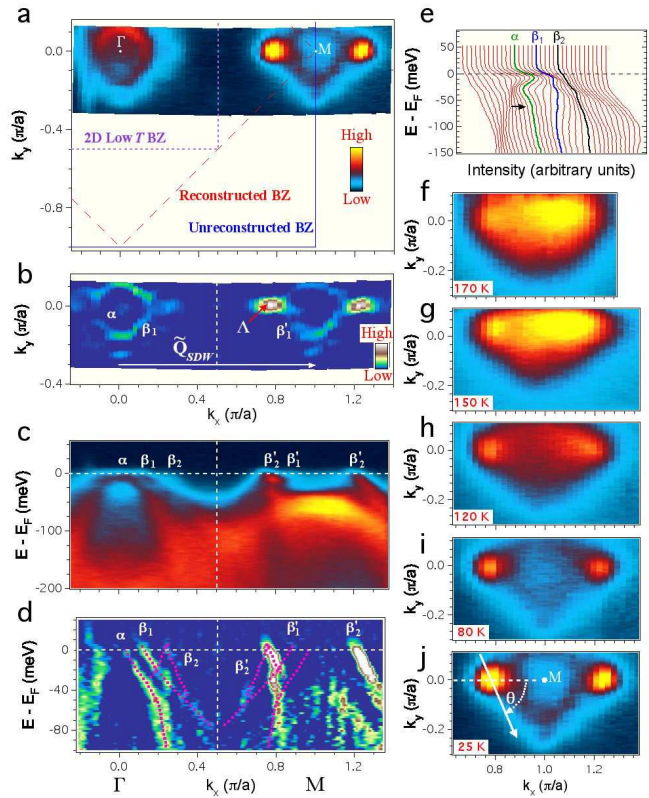}
\caption{\label{Fig_map}(Color online) (a) FS mapping (25 K) obtained by integrating the photonemission intensity in a 20 meV window centered at $E_F$. The FS is described in terms of the unreconstructed BZ. (b) Corresponding second derivative intensity plot. $\widetilde{Q}_{SDW}$ is the in-plane projection of the SDW wave vector. (c) and (d) ARPES intensity plot and corresponding second momentum derivative intensity plot, respectively, along the $\Gamma$(0, 0)-M($\pi$, 0) symmetry line. The dotted lines are guides for the eye. The vertical dashed lines in (b)-(d) indicate the 2D low $T$ BZ. (e) EDCs around the $\Gamma$ point. The $\alpha$ band is touching $E_F$ at the $\Gamma$ point. The arrow indicates the bottom of an electronlike dispersion that hybridizes with the $\alpha$ and $\beta_1$ bands. (f)-(j) $T$ evolution of the FS around the M point. The definition of the $\theta$ angle is given in panel (j).} 
\end{center}
\end{figure}

In Fig. \ref{Fig_map}(a), we show a FS mapping obtained at 25 K. Two regions of similar size with high photoemission intensity are centered around the zone center ($\Gamma$ point) and the M($\pi$, 0) point [($\pi$, $\pi$) in the reconstructed BZ], respectively. The most remarkable features and main focus of this Letter is a very bright spot observed around (0.75$\pi$, 0) on the $\Gamma$-M symmetry line. The second derivative intensity plot of the FS shown in Fig. \ref{Fig_map}(b) suggests that the FS is folded across the 2D low temperature ($T$) BZ and that the $\Gamma$- and M-centered FSs are replica connected by the $\widetilde{Q}_{SDW}$=($\pi$, 0) wave vector [($\pi$, $\pi$) in the reconstructed BZ description]. This is quite clear for the $\beta_1$ FS and its reflection $\beta_1^{\prime}$, as well as for the bright spot and its equivalent symmetry points ($\pm$0.25$\pi$, 0). We note that although the FS pattern is 4-fold symmetric around the $\Gamma$ and M points, our results do not allow us to distinguish between an intrinsic 4-fold symmetry and the superimposition of twin domains. 

The intensity plot of the $\Gamma$-M cut is given in Fig. \ref{Fig_map}(c) and the corresponding second momentum derivative intensity plot (2MIP) is displayed in Fig. \ref{Fig_map}(d). In addition to the $\alpha$ and $\beta_1$ bands expected from a comparison with optimally doped samples \cite{Ding_EPL}, an extra holelike dispersion crossing $E_F$ ($\beta_2$) is observed around the $\Gamma$ point, in agreement with previous reports \cite{Yang_LX, LiuPRL2009}. This band is also folded across the 2D low $T$ BZ boundary. Interestingly, it crosses $E_F$ at $\Lambda$=(0.75$\pi$, 0), which corresponds to the bright spot. The $T$ evolution of the FS, displayed in Figs. \ref{Fig_map}(f)-\ref{Fig_map}(j), indicates that while the bright spot at the $\Lambda$ point is clearly visible up to 120 K, it becomes hard to identify at 150 K and above, reinforcing our assumption that the spot at the $\Lambda$ point is formed, at least partly, by an extra band that appears below $T_{SDW}$ (138 K). Interestingly, an electronlike band crosses $E_F$ at the same point. As expected, the energy distribution curves (EDCs) given in Fig. \ref{Fig_map}(e) show that this band is folded to the $\Gamma$ point. Moreover, it hybridizes with both the $\alpha$ and $\beta_1$ bands, opening energy gaps below $E_F$. 

\begin{figure}[htbp]
\begin{center}
\includegraphics[width=8.8cm]{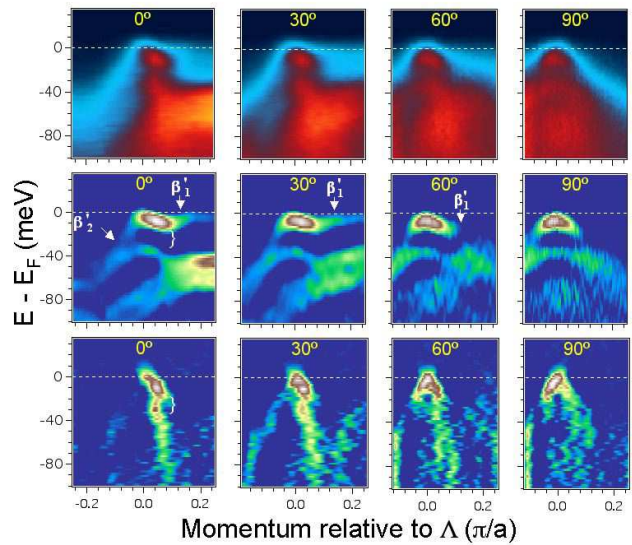}
\caption{\label{Fig_25K}(Color online) ARPES intensity plots (25 K) for cuts going trough $\Lambda$(075$\pi$, 0) are displayed in the top row and indexed after the $\theta$ angle defined in Fig.\ref{Fig_map}(j). The corresponding second energy derivative and second momentum derivative intensity plots are shown in the second and third rows, respectively. The direction of the straight arrow in Fig.\ref{Fig_map}(j) defines the momentum increase, which is towards M for $\theta=0$.}
\end{center}
\end{figure}

One question immediately follows: does that electronlike band hybridize with the $\beta_2^{\prime}$ band around the $\Lambda$ point as well? To answer this question and characterize the $\Lambda$ point further, we plot in the top row of Fig. \ref{Fig_25K} cuts \emph{going through} the $\Lambda$ point and indexed according to the $\theta$ angle defined in Fig. \ref{Fig_map}(j). The corresponding second energy derivative intensity plots (2EIPs) and 2MIPs are displayed in the second and third rows of Fig. \ref{Fig_25K}, respectively. The figures exhibit a complicated hybridization pattern. For instance, the hybridization between the electronlike band and the $\beta_1^{\prime}$ at $\theta$=0 manifests itself by a gap in the 2EIP in the $\sim$20-30 meV binding energy range and close to a vertical section of the dispersion for the same energy range in the 2MIP. 

\begin{figure}[htbp]
\begin{center}
\includegraphics[width=8.8cm]{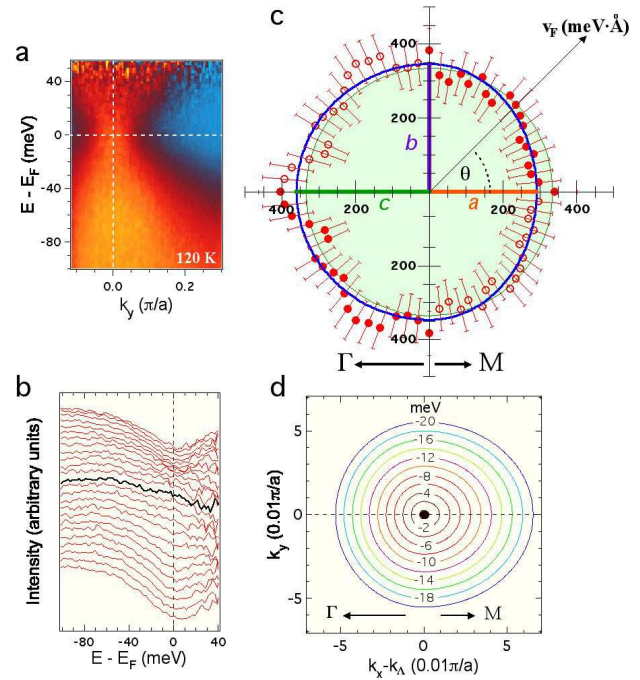}
\caption{\label{Fig_VF}(Color online) (a) ARPES intensity spectra of BaFe$_2$As$_2$ at the $\Lambda$ point ($\theta=90^{\circ}$) recorded at 120 K, after division by the Fermi-Dirac function convoluted with the instrumental resolution function. The corresponding EDCs are given in panel (b), where the bold EDC refers to the $\Lambda$ point. (c) Polar representation of $v_F$ around the $\Lambda$ point (25 K). Open and close circles represent data measured and data obtained by reflection with respect to the $\Gamma$-M symmetry line, respectively. The large filled circle represents the average value of $v_F$ while the thick line is a fit of the data to the two-ellipse model described in the text, with parameters $a$, $b$ and $c$. (d) Contour plot of the electronic dispersion below $E_F$ around the $\Lambda$ point, as calculated from our model. The small filled circle represents the FS associated with the cone.}
\end{center}
\end{figure}

Surprisingly, the portion of the band structure around the $\Lambda$ point, from $E_F$ to a binding energy of about 20 meV, seems not to change much with $\theta$: an electronlike band and a holelike band intersect around $E_F$ and their angular symmetry defines a cone structure, well illustrated by the angular evolution of the 2MIPs, which are not distorted by the Fermi edge. By dividing a spectra recorded at 120 K ($<T_{SDW}$) by the Fermi-Dirac function convoluted with the instrumental resolution [Fig. \ref{Fig_VF}(a)], we can partly access the electronic band structure above $E_F$. From the EDCs around the $\Lambda$ point, which are given in Fig. \ref{Fig_VF}(b), we conclude to the absence of a hybridization gap at the $\Lambda$ point, at least within our resolution. Hereafter, we thus refer to this structure as a Dirac cone, as observed in graphene \cite{Novoselov}.   

Unlike graphene, for which the Dirac cone is isotropic and located at a high symmetry point, the $\Lambda$ point in BaFe$_2$As$_2$ is not a high-symmetry point. To investigate the possible anisotropy of the band structure around $\Lambda$, we extracted the Fermi velocity ($v_F$) as a function of $\theta$ by assuming a linear dispersion in the vicinity of $E_F$. The data are reported on a polar plot in Fig. \ref{Fig_VF}(c). The size of the plain circle represents the 330 $\pm$ 60 meV$\cdot$\AA\xspace average of $v_F$. Although this circle is consistent with the data within uncertainties, a better agreement is achieved if we assume a small anisotropy. Since symmetry across the $\Gamma$-M axis is the only constraint imposed at the $\Lambda$ point, we fitted the data with two ellipses defined by half-axes ($a$, $b$) and ($c$, $b$), respectively, the $b$-axis being perpendicular to the $\Gamma$-M direction, \emph{i.e.} for $\theta=\pm90^{\circ}$. As illustrated in Fig. \ref{Fig_VF}(c), we obtained $v_F$($a$) = 290$\pm60$ meV$\cdot$\AA, $v_F$($b$) = 350$\pm60$ meV$\cdot$\AA\xspace and $v_F$($c$) = 360$\pm60$ meV$\cdot$\AA, respectively. 

\begin{figure}[htbp]
\begin{center}
\includegraphics[width=8.8cm]{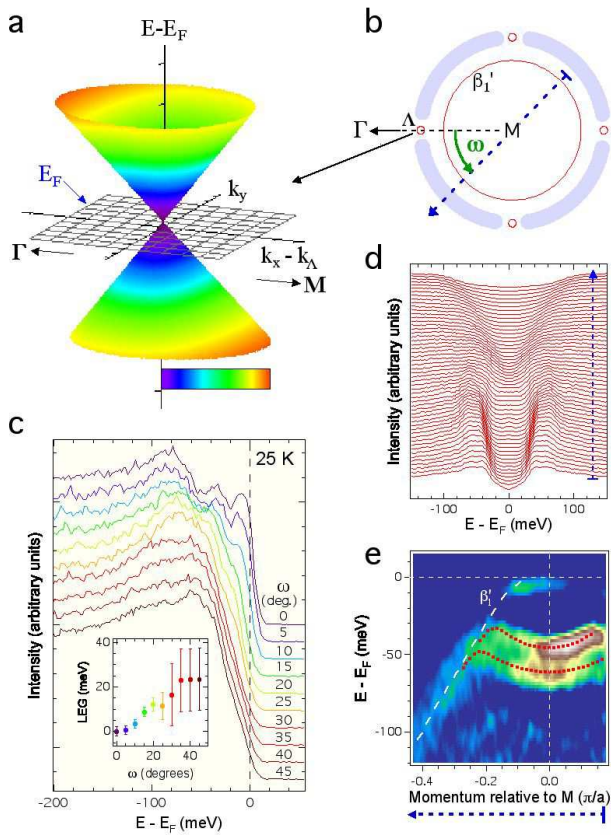}
\caption{\label{Fig_gap}(Color online) (a) 3D representation of the Dirac cone at the $\Lambda$ point. The color scale indicates the distance from the $\Lambda$ point. (b) Schematic representation of the FS around the M point. The folded $\alpha$ band, which is barely touching $E_F$, is not indicated. Shaded areas indicate gapped regions and the dashed arrow indicates the orientation of the ARPES cut associated with (d) and (e). (c) Minimum gap locus EDCs (25 K) as a function of the angle $\omega$ defined in panel (b). The inset shows the LEG as a function of $\omega$ after a 3.5 meV shift (see the text). (d) Symmetrized EDCs (25 K) along the cut indicated in panel (b). (e) 2EIP (25 K) along the cut indicated in panel (b). The vertical dashed line indicates the M point.}
\end{center}
\end{figure}

Our analysis reveals that the apex of the cone is located 1$\pm$5 meV above $E_F$, which corresponds to $E_F$ within uncertainties. Fig. \ref{Fig_VF}(d) shows a contour plot of the cone dispersion below $E_F$ reconstructed by assuming a linear dispersion, a cone apex located 1 meV above $E_F$ and $v_F$ values derived from the $a$, $b$ and $c$ parameters. It can be used to determine the evolution of the cone FS size as a function of the chemical potential position. With the cone apex located 1 meV above $E_F$, the FS size, which corresponds to the small filled surface in Fig. \ref{Fig_VF}(d), is only 1$\times$10$^{-3}\%$ of the reconstructed BZ, a value smaller than the estimation obtained from QOE in \SFA\xspace\cite{Sebastian} and \BFA\xspace\cite{Analytis}. However, the FS size is very sensitive to the position of $E_F$. While a downward chemical potential shift of 8 meV would lead to an area of 1$\%$ of the reconstructed BZ similar to the values reported by QOE \cite{Sebastian, Analytis}, an upward shift of only slightly more than 1 meV would switch the FS from holelike to electronlike. In Fig. \ref{Fig_gap}(a), we plot a three-dimensional representation of the anisotropic Dirac cone in the vicinity of $E_F$ calculated from our model.

What physical mechanism can generate such a topological anomaly in the electronic band structure and the FS of BaFe$_2$As$_2$? Within the framework of SDW mean-field theory of both two-band and five-band tight-binding models, Ran \emph{et al.} demonstrated that nodes in the SDW gap function of undoped FeAs-based superconductors must exist due to the occurrence of Dirac cones forming as a consequence of the symmetry-enforced degeneracy at the $\Gamma$ and M high-symmetry points, even in the presence of perfect nesting \cite{Ran}. Interestingly, these nodes are located in the same region as the cone that we observe. The gap function is predicted to vary away from the cone. We investigated this scenario by measuring the leading edge gap (LEG) of EDCs at the minimum gap locus (MGL) on cuts around the M point. For this, we define the angle $\omega$ in Fig. \ref{Fig_gap}(b) and display the results in Fig. \ref{Fig_gap}(c). As $\omega$ increases, the LEG shifts towards high binding energy. The peak is also suppressed and the position of the MGL is hard to define above 25 degrees, leading to large uncertainties in the value of the LEG, which is given as a function of $\omega$ in the inset of Fig. \ref{Fig_gap}(c). We also note that the LEG at $\omega$=0 is located 3.5 meV above $E_F$ and thus we shifted LEG ($\omega$) by that value. 

Even though the MGL of the bands forming the Dirac cone is hard to determine for large $\omega$ angles, both the symmetrized EDCs [Fig. \ref{Fig_gap}(d)] and the 2EIP [Fig. \ref{Fig_gap}(e)] corresponding to the cut  indicated by a dashed arrow in Fig. \ref{Fig_gap}(b), suggest that they are gapped. In fact, as supported by the 2EIP given in Fig. \ref{Fig_gap}(e), two electronlike bands exhibit gaps of $\sim$ 50 and $\sim$30 meV. Since the apex of the cone is located approximately at $E_F$, it is a good first approximation to assume that SDW gaps will be centered near $E_F$ as well. Our results would correspond to full SDW gaps of $\sim$ 100 and $\sim$60 meV, respectively. Although further studies would be necessary to conclude that these gap values correspond to optical gaps, it is remarkable that they agree reasonably well with the 110 and 45 meV gaps reported from optical measurements on BaFe$_2$As$_2$ single crystals \cite{WZ_Hu}. 

We caution that we did not consider band dispersion along the $k_z$ direction \cite{LiuPRL2009,Vilmercati, Malaeb}. Although a strong $k_z$ dependence is found for the $\alpha$ band, Liu \emph{et al.} reported only little variations for the $\beta_1$ and $\beta_2$ bands \cite{LiuPRL2009}. The latter, which coincides with the $\Lambda$ point, is found around both the $\Gamma$ and M points, confirming band folding. Due to the sensitivity of the Dirac cone apex position to the precise electronlike and holelike dispersions generating the cone, it is likely that even a very small wiggling in the $k_z$ dispersion can lead to a shift of the cone apex as a function of $k_z$ and produce the series of very small FSs suggested from QOE \cite{Sebastian,Analytis}, whose location in the momentum space is now determined. Even though the failure of QOE to detect the large holelike $\beta_1$ FS pocket observed by ARPES and predicted by calculations remains puzzling, our observation of tiny FSs in the parent compound of an iron-pnictide superconductor derived from a fundamental topological anomaly, \emph{i.e.} a Dirac cone dispersion, is of crucial importance to understand the unconventional electronic properties of these materials and allows us to describe them as nodal SDW materials.

We thank Z. Wang and S. Zhou for valuable discussions. This work was supported by grants from JSPS, JST-TRIP, JST-CREST, MEXT of Japan, the Chinese Academy of Sciences, Ministry of Science and Technology of China, and NSF of US.

\bibliography{biblio_ens}

\end{document}